\documentclass[10pt]{article}

\usepackage[utf8]{inputenc}
\usepackage{enumitem}
\usepackage{graphicx}

\title{kafe2 - a Modern Tool for Model Fitting in Physics Lab Courses}

\author{Johannes Gäßler$^{(1)}$,
  Günter Quast$^{(1)}$,
  Daniel Savoiu$^{(2)}$,
  Cedric Verstege$^{(1)}$ \\
  $^{(1)}$ Karlsruhe Institute of Technology (KIT), 
  $^{(2)}$ Hamburg University \\
	}

\date{\today}

\begin{document}

\maketitle

\begin{abstract}
  Fitting models to measured data is one of the standard tasks in the natural sciences,
  typically addressed early on in physics education in the context of
  laboratory courses, in which statistical methods play a central role in analysing and interpreting experimental results.
  The increased emphasis placed on such methods in modern school curricula, together with the
  availability of powerful free and open-source software tools geared towards scientific data analysis, form an excellent premise for the development of new teaching concepts for these methods at the university level.
  In this article, we present {\it kafe2}, a new tool developed at the Faculty of Physics at the Karlsruhe 
  Institute of Technology, which has been used in physics laboratory courses for
  several years.
  Written in the {\it Python} programming language and making extensive use of established numerical and optimization libraries, {\it kafe2} provides simple but powerful interfaces for numerically fitting model functions to data.
  The tools provided allow for fine-grained control over many aspects of the fitting procedure, including the
  specification of the input data and of arbitrarily complex model functions,
  %using native {\it Python} syntax,
  the construction of complex uncertainty models, %taking multiple uncertainty sources and %correlations into account,
  and the visualization of
  the resulting confidence intervals of the model parameters.
\end{abstract}

\section{Model fitting in lab courses: pitfalls and limitations}
 \label{sec_intro}

A primary goal of data analysis in physics is to check whether an assumed model adequately describes the experimental data within the scope of their uncertainties. A regression
procedure is used to infer the causal relationships between independent and dependent variables, the abscissa values $x_i$ and the ordinate values $y_i$. The dependency is typically represented by a functional relation $y_i = f(x_i; \{p_k\})$ with a set of model parameters $\{p_k\}$. 
Both the $y_i$ and the $x_i$ are the results of measurements and hence are affected
by unavoidable measurement uncertainties that must be taken into account when determining
confidence intervals for the model parameters. It has become a common standard in the natural sciences to determine confidence intervals for the parameter values based on a
maximum-likelihood estimation (MLE). In practice, one typically uses the negative
logarithm of the likelihood function (NLL), which is minimised by means of numerical
methods.

In the basic laboratory courses, parameter estimation is traditionally based on the  
method of least squares (LSQ), which in many cases is equivalent to the NLL method. 
For example, fitting a straight line to data affected by static Gaussian uncertainties
on the ordinate values $y_i$ is an analytically solvable problem and the necessary
calculations can be carried out by means of a pocket calculator or a simple spreadsheet application. For simplicity, non-linear problems are often converted to linear ones by 
transforming the $y$ values. However, the Gaussian nature of the uncertainties is lost in this process, which is why the transformed problem only gives accurate results if the uncertainties are sufficiently small. 
Simple error propagation is often used to transform the fitted parameters like the
slope or $y$ intercept to the model parameters of interest. As a result, the 
parameter uncertainties obtained in this way are no longer Gaussian, and it is
non-trivial to determine sensible confidence intervals. When the independent
variables are also affected by measurement uncertainties, or when  relative 
uncertainties with respect to the true values are present, this traditional approach
reaches its limits.

 Such simple analytical methods are very limited in their general applicability to real-world problems. For instance, the presence of uncertainties on the abscissa 
 values in addition to those on the ordinate values, as they would be encountered when
 measuring a simple current-voltage characteristic of an electronic component, already results in a problem that can generally no longer be solved analytically. Unfortunately, most of the widely used numerical tools for parameter estimation provide only limited support for such non-linear problems, even though they occur frequently in practice.
 
 Prior to drawing any conclusions on model parameters, the most important task in physics
 is to check the validity of a model hypothesis assumed to describe a set of measurement data. The full task therefore consists of the following steps:
 \begin{enumerate}
 \item carefully quantifying the uncertainties of all input measurements;
 \item defining a model hypothesis for the measured data;
 \item testing whether the measurements are compatible with the model hypothesis;
 \item if so, determining the model parameters, e.g.\ the slope and intercept of a 
 straight line.
 \end{enumerate}

 It is precisely the hypothesis testing in step 3 that drives progress in scientific
 understanding by providing the relation between measurements and theoretical 
 models. A simple example for such a hypothesis test is the $\chi^2$ test, which
 results directly from the frequently used method of least squares. A test for the
 validity of the model assumption should definitely be performed before drawing any 
 conclusions from the values and uncertainties of the fitted model parameters.
 Unfortunately, many of the common tools come with a default setting that assumes the
 correctness of the given parameterization and scales the parameter uncertainties
 so that the model perfectly describes the data within the parameter uncertainties 
 so determined. This behaviour can usually be switched off by choosing appropriate
 options, but in practice this is often neglected. 

 The parameters of a complex model can be highly correlated.
 In this case it is no longer sufficient to consider the distribution of just a single parameter.
 Instead the shared distribution of multiple parameters needs to be considered, for example when conducting hypothesis testing.
 Correlations between model parameters can also be problematic for numerical optimisation.
 For this reason strategies for choosing appropriate 
 parameterizations to reduce these correlations also need to be addressed. 
 It is then indispensable that the correlations are also shown 
 when presenting the fit results.
 Ideally the correlation between two model parameters can be expressed with a simple correlation coefficient.
 However, if non-linear models are fitted, this is often not sufficient for the 
 evaluation of the result; in this case the corresponding confidence regions 
 should be determined and presented as contour graphs for pairs of parameters. 

 The distinction between "statistical" and "systematic" uncertainties leads to 
 many misunderstandings among students. For example, the systematic error of 
 a measuring instrument becomes a statistical one if several measuring instruments 
 of the same type are used. A much more suitable approach is to differentiate whether
 uncertainties affect all measured values or groups of measured values equally
 or whether they are independent. This approach, however, requires dealing with 
 the covariance matrix of measurement uncertainties and constructing it 
 from the individual uncertainties on a problem-by-problem basis. Unfortunately, 
 there are very few simple tools that allow a full covariance matrix to be considered
 in the fitting process. None of the tools commonly used in physics lab courses allow
 students to take into account correlated uncertainties of both the abscissa and ordinate
 values.

 While there are tools with which the requirements listed here could be partially 
 implemented, these do not provide the simplicity needed for undergraduate physics
 education. To address the issues described above it became necessary to write a tool
 of our own, {\it kafe2}, an open-source Python package designed to provide a flexible 
 Python interface for the estimation of model parameters from measured data.

\section{The Open-Source Python Package {\it kafe2}}
 \label{sec_kafe2}
 
 After designing a prototype package, {\it kafe}, development continued with 
 {\it kafe2}\,\cite{kafe2}. 
Said tools make use of contemporary methods for visualization and evaluation of 
measurement data and are based on freely available, open-source software that is 
also used in scientific contexts and thus relevant for later professional practice.

 \subsection{Implementation} \label{ssec_implementation}  

Due to its widespread use in scientific communities as well as in the burgeoning professional field of 
data science, \emph{Python}\,\cite{python} was chosen as the programming language for {\it kafe2}.
As a high-level language with particular emphasis on clear and intuitive syntax, and convenient features such as dynamic typing and garbage collection, Python provides a beginner-friendly programming environment, requiring comparatively little knowledge to use effectively.
This is particularly advantageous in a field such as physics, which requires a very broad foundation of
knowledge prior to specialization and thus allows only a limited percentage of the curriculum to be assigned to programming.

However, an inevitable downside of Python is that it is comparatively slow compared to 
low-level languages such as C/C++.
Fortunately, the Python ecosystem features many libraries for numerical computation, most notably the \emph{NumPy}\,\cite{numpy} library, which makes use of fast underlying C implementations for running such computations efficiently.
Optimal use of these libraries does require some detailed knowledge, but in the case of 
{\it kafe2} this generally only applies to development rather than use.

The numerical optimization algorithms necessary for finding the global minimum of the negative 
log-likelihood function are provided either by the library \emph{SciPy}\,\cite{scipy} ("scientific Python") 
or by the Python package \emph{iminuit}\,\cite{iminuit}, which is based on the \emph{MINUIT} package developed at 
CERN, the European Organization for Nuclear Research.

The graphics library \emph{matplotlib}\,\cite{matplotlib} is used as the graphical backend for publication 
quality visualizations of input data and results.
{\it kafe2} itself is implemented in pure Python and can therefore run on all common 
platforms with a Python environment, including ARM-based platforms such as the Raspberry Pi.

\subsection{Special Features of {\it kafe2}} \label{ssec_features}

The {\it kafe2} package provides numerous unique features that are listed in the following.

To represent complex uncertainty models with absolute and/or relative, independent and/or correlated uncertainties
on abscissa and/or ordinate values, {\it kafe2} uses two covariance matrices collecting the uncertainty components
on the abscissa and ordinate directions. As described below, these are combined into a single overall covariance matrix used in the numerical regression.
Users can specify multiple uncertainties of the same kind which are then automatically combined into the corresponding covariance matrices. 
If relative uncertainties are specified, they most often refer to the true
values of the input data, which are only (approximately) known after the regression. {\it kafe2} is able to handle 
such cases by dynamically calculating the uncertainties as relative to the model;
optionally and for compatibility with other tools it is possible to use relative uncertainties 
with respect to the measured data. Correlations between uncertainty components or between individual uncertainties can
be specified as a correlation coefficient or as a complete covariance matrix. 

In addition to pre-defined, rather simple model functions, arbitrarily complex user-defined functions can be
easily implemented using native Python syntax. It is also possible to simultaneously fit several model functions 
with shared parameters to data; this feature is very useful if auxiliary or external measurements of model parameters
need to be considered. Fixing the values of one or more model parameters or constraining them within given external uncertainties
is also supported. 

The numerical fitting procedure in {\it kafe2} relies on the NLL method, but falls back to the simple
LSQ method if equivalent to NLL in order to avoid computationally expensive operations.  Distributions can be fitted to
binned data (histograms) using a Poisson likelihood; optionally and for compatibility with
other packages the LSQ method can alternatively be chosen. {\it kafe2} also supports fits
with user-defined maximum-likelihood functions. 

Parameter uncertainties are determined in the traditional way relying on the 
the Cramér-Rao-Fréchet bound, i.e.\ from the second derivatives of the LSQ or NLL function at the
minimum. To deal with non-linear cases and correlated parameters, a likelihood-based method known as 
"profile likelihood" is also optionally offered for determining parameter confidence intervals. 
In multi-parameter problems, this is also useful to account for the influence of nuisance 
 parameters when determining individual parameter uncertainties.

Fit results are provided in graphical form, but output in text form is also possible. Access
functions exist to extract results for further processing with Python code. If meaningful, {\it kafe2}
provides metrics to judge the goodness of a fit result.  The output also includes the parameter correlations,  
as is standard for most other fitting tools. The graphical output shows the data and the best-fit model 
function with an optional representation of the model uncertainty as a shaded band. 
An optional graphical representation of the profile likelihood for each parameter and of confidence 
regions of pairs of parameters as contours are also provided. 

More details on the mathematical procedures and a concrete application example are given below.

\subsection{Mathematical procedures} \label{ssec_mathematics}

The estimation of model parameters and their uncertainties in {\it kafe2}
is based on the method of maximum-likelihood estimation. Given a set of $N$ measurements 
$y_i$ with corresponding abscissa values $x_i$, as well as a model function 
$f({x_i}, \{p_k\})$ depending on a set of parameters $\{p_k\}$, the most commonly assumed 
data distribution is the $N$-dimensional Gaussian distribution $\cal N$ with the uncertainties 
on the ordinate values being described by the covariance 
matrix $\bf V$:
%\[
%\mathcal{N}\left( y_i, f(x_i, \{p_k\}), {\bf V} \right) \,=\, 
%     \displaystyle \frac {1} {\sqrt{(2\pi)^N \det {\bf V}}} \exp \left(-\frac{1}{2}
%     \sum_{i,j=1}^N    \left( y_i - f(x_i, \{p\}) \right) 
%     \left( {\bf V}^{-1} \right)_{ij}  \left( y_j - f(x_j, \{p\}) \right) \right)\,.
%\]
\[
\mathcal{N}\left( \{y_i\}, f(\{x_i\}, \{p_k\}), {\bf V} \right) \,=\, 
     \displaystyle \frac {1} {\sqrt{(2\pi)^N \det {\bf V}}} \exp \left(-\frac{1}{2}
     \sum_{i,j=1}^N    \left( y_i - f(x_i, \{p_k\}) \right) 
     \left( {\bf V}^{-1} \right)_{ij}  \left( y_j - f(x_j, \{p_k\}) \right) \right)\,.
\]

The cost function ${\cal{C}}$ to be minimized is defined as twice the negative logarithm
of the likelihood ${\cal{L}}$ of $\cal N$, ${\cal{C}} = -2 \ln {\cal{L}}$.
Terms that are independent of the parameters are typically neglected in the definition of
the likelihood since they are irrelevant for the location and shape of the global cost function minimum.

The total uncertainties of the measurement data are described by the covariance matrix 
${\bf V}$. The uncertainties of the measurement points in the abscissa direction 
are described by a covariance matrix with elements 
${\bf V}^x_{ij}$. Because the Gaussian likelihood is only defined for ordinate uncertainties
{\it kafe2} handles abscissa uncertainties by transforming them into
ordinate uncertainties. This is achieved by multiplying them with the first derivatives 
of the model function.
The resulting uncertainties are then simply added to the covariance matrix elements ${\bf V}^y_{ij}$ of the ordinate uncertainties. 
The elements of the total covariance matrix used in the fit are thus

\[
{\bf V}_{ij} = {\bf V}^y_{ij} + {\bf V}^x_{ij} \cdot f'(x_i) \cdot f'(x_j)\,.
\]

This approach assumes that, near the global cost function 
minimum, the fitted model function can be sufficiently well approximated by a straight line in the vicinity of the support points $x_i$, i.e.\ that second-order derivatives are negligible on the scale of the abscissa uncertainties $\Delta x_i$.

For uncertainties that are independent of the parameter values,  ${\cal{C}}$ is equivalent
to the method of least squares, which is classically  used as the cost function in most 
fitting algorithms, 
\[
{\cal{C}}(\{y_i\}, {\bf V}^y, f(\{x_i\},\{p_k\}) ) \,=\,  
 \displaystyle\sum_{i,j}    \left( y_i - f(x_i, \{p_k\}) \right) 
     \left( \big({{\bf V}^y}\big)^{-1} \right)_{ij}  \left( y_j - f(x_j, \{p_k\}) \right)\,.
\]

The expected value of ${\cal{C}}$ for the best parameter values estimated from a 
hypothetical set of data follows a $\chi^2$-distribution with a number of 
degrees of freedom (NDF) corresponding to the difference between the number of 
data points and the number of free model parameters. The ratio $\mathcal{C}/\mathrm{NDF}$,
often called the "reduced $\chi^2$", has an expectation value of unity for a model that
accurately describes the data within their uncertainties. 
The "$\chi^2$ probability" $p = 1 - q$ is determined 
from the quantile $q$ of the $\chi^2$ distribution for the observed
minimum and  can be used to quantify the goodness-of-fit.
If, however, the uncertainties do depend on the parameters, as is for instance the case for relative ordinate
uncertainties or uncertainties on the abscissa values, then the method of least squares
is no longer equivalent to the method of maximum likelihood.
This is because the normalization term of the multivariate Gaussian, which is a function of
$\det\left(\bf V(\{p_k\})\right)$, must also be taken into account.
While a parameter-dependent covariance matrix needs to be recomputed on every parameter change, thus adding to the computational cost of the minimization, it remains feasible on modern hardware, particularly if the inversion of the covariance matrix 
$\bf V$ is replaced by a Cholesky decomposition ${\bf L} {\bf L}^T = {\bf V}$.
In addition, the Cholesky decomposition also enables the 
quick computation of the covariance matrix determinant by exploiting that 
$\bf L$ is a triangular matrix: 
$\det\left({\bf V}\right) = 
  \det\left({\bf L}\right) \cdot \det\left({\bf L}^T\right) =
  2 \det\left({\bf L}\right)$.

Following common standards based on the Cramér-Rao-Fréchet bound, an estimate of the covariance matrix of the parameter uncertainties is determined from the inverse Hessian matrix, containing the second-order derivatives of 
${\cal{C}}(\{y_i\}, V, f(\{x_i\};\{p_k\}))$ with respect to the parameters $p_k$.
In general, however, this approach is not sufficient for reliably determining 
the confidence intervals of the fitted parameters. For this reason, {\it kafe2} also 
implements the \emph{profile likelihood} method.
The so-called profile as a function of a subset of the parameters is determined by
keeping the values of these parameters fixed and minimizing ${\cal{C}}$ with 
respect to all other parameters. The boundaries of a confidence interval are determined by 
finding the locus of points for which the profile likelihood increases by a value
$\Delta\mathcal{C}$ with respect to the global minimum.
In the case of two parameters,
this amounts to finding the intersection of the two-dimensional profile likelihood function 
with a horizontal plane above the global cost function minimum.

The value of $\Delta{\cal{C}}$ determines the confidence level of the confidence interval/region.
For example, for a one-dimensional confidence interval $\Delta{\cal{C}} = 1$ corresponds 
to the $1\,\sigma$ interval of a Gaussian distribution with a confidence level of 68.3\%, 
where $\sigma$ is the standard deviation of the one-dimensional Gaussian distribution.
In the two-dimensional case (as in the profile likelihood of two parameters) the 
intersection of a two-dimensional profile likelihood with a horizontal plane is a 
one-dimensional curve. This curve corresponds to a confidence contour, which is an
ellipse in case of Gaussian uncertainties\footnote{
It should be noted that the uncertainties on the fitted model parameters can often 
be made more Gaussian-like by changing the parameterization of the model function.}.
If the  contour deviates strongly from the elliptical form, is no longer possible to 
correctly specify a confidence region of the two parameters by providing 
a simple correlation coefficient. In this case, the contour plots of the 
affected pairs of parameters as well as the one-dimensional confidence 
intervals should be documented.

The profile likelihood method is important because non-linear problems are very common in
practice. Even a simple linear regression with additional uncertainties on the abscissa 
values leads to a non-linear problem for which no general analytical solution exists.
Relative uncertainties on the ordinate values introduce similar issues.
They should be specified relative to the model (as a stand-in for the unknown
"true values") to avoid the bias which otherwise would give too much weight to
data points with downward fluctuations of the measured values and hence the 
assigned uncertainties. 
When the model function itself is nonlinear in the parameters or when abscissa uncertainties 
or relative ordinate uncertainties are used, the shape of the covariance contours should always 
be inspected. Using appropriate tools, the associated calculations usually only take on the 
order of seconds on consumer hardware.
An example with {\it kafe2} is shown below. 

In addition to the Gaussian-based cost functions discussed until now, {\it kafe2}
also offers other cost functions, for example a cost function based on the Poisson distribution 
for fitting models to frequency distributions (histograms). The specification of user-defined cost 
functions is also possible; in order to benefit from the statistical interpretation of the 
fit results provided by {\it kafe2}, such cost functions must be based on valid likelihood
functions.

Often it is desirable to show the uncertainty of the fitted model function. 
An approximate confidence region $\Delta f(x; p)$ can be derived from the 
covariance matrix, ${\bf V}^p$, of the model parameters and the gradient, 
$\nabla_p f(\{x_i\}; \{p_k\})$, of the model function with respect to the 
parameters by (linear) propagation of the uncertainties:
$$ \Delta f(x_i, \{p_k\}) = 
 \Big( \nabla_p f(x_i; \{p_k\}) \Big)^T {\bf V}^p \  \nabla_p f(x_i; \{p_k\})\,.$$

\section{User interface and applications} \label{sec_userinterface}  

To mitigate the inherent trade-off between usability and complex
functionality {\it kafe2} offers the user several interfaces for fitting:
\begin{itemize}
 \item  For those without programming knowledge, or for the convenience of
 advanced users, a command line interface “\emph{kafe2go}” is provided. 
 Users only need to supply a configuration file in the format of the widely
 used data serialization language YAML\,\cite{yaml}, which specifies the data,
 the model function and fitting options. 
 \item  More flexibility is offered to slightly advanced users with
 basic programming knowledge; they can call the Python interface of built-in 
 {\it kafe2} pipelines as part of a larger Python script.
 \item Advanced users have the possibility to construct custom pipelines by 
 directly instantiating the {\it kafe2} objects that represent input data, 
 model functions, fits or graphical output.
\end{itemize}

As an example, the stand-alone application
{\it kafe2go} can be used via the following command line call:

 \verb| kafe2go <name>.yaml|. 

\begin{figure}[hbtp] \begin{center}
\mbox{\includegraphics[width=0.9\linewidth] {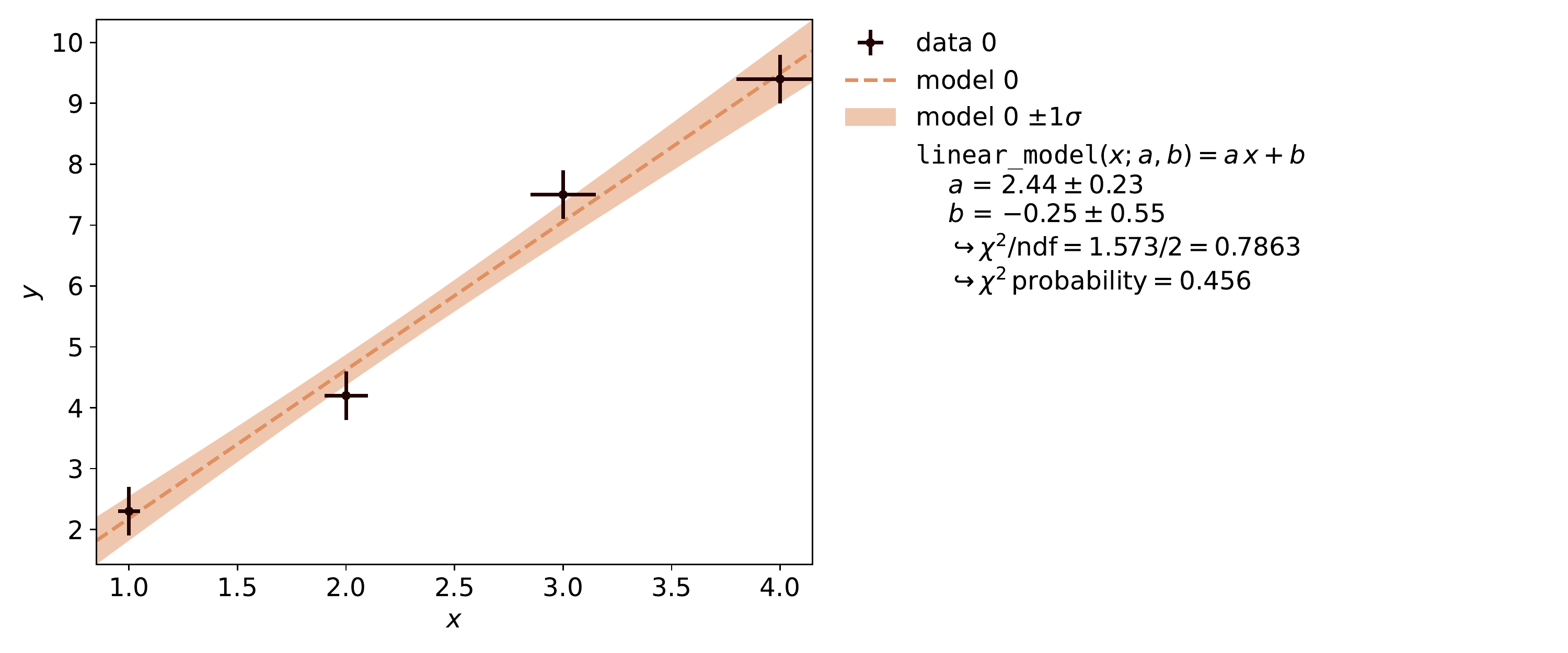}}
\vspace {-1pc} \caption[Example of a fit with kafe2] 
{Graphical output of {\it kafe2}. The shaded region around the fitted 
 straight line shows the 68\% confidence interval of the model function.} 
\label{fig:fitExample} 
\end{center}\end{figure}

Figure~\ref{fig:fitExample} illustrates the result of fitting a straight line to 
data points with uncertainties in both the ordinate and abscissa directions.
The input file contains only the following lines: 
\pagebreak[2]

 \begin{verbatim}
 # input data straight line fitting
 x_data: [1.0, 2.0, 3.0, 4.0]
 x_errors: 5% 
 y_data: [2.3, 4.2, 7.5, 9.4] 
 y_errors: 0.4 
 \end{verbatim}
 
Note: if no model function is specified, {\it kafe2go} defaults to a first-degree 
polynomial as the model. 

When using {\it kafe2} as a Python library, a Jupyter\,\cite{jupyter} environment
is very convenient. One advantage is that this allows users to combine text, 
program code and outputs in a so-called "Jupyter notebook".
A second advantage is that it also enables teaching staff to provide the entire Python 
environment for running {\it kafe2} as a service to students;
the students themselves only need to connect to a Jupyter server via web browser without 
the need to spend time setting up a Python environment themselves.

To use the built-in {\it kafe2} pipelines for model fitting, users simply need to
call the corresponding functions.
The creation of custom pipelines, however, requires a more advanced understanding
of object-oriented programming. First, a data container object is instantiated 
from numerical input data and their uncertainties.
Then a fit object is created from a data container object and a model function.
After possibly constraining or limiting parameters via dedicated method calls,
the method \verb|do_fit()| is called to perform the numerical optimization of 
the cost function. The fit results are then accessed via the corresponding
attributes of the fit object or printed to console by calling the
\verb|report()| method. In order to create a graphical representation of 
the fit results a plot object can be instantiated from the fit object.
A complete code example, which corresponds exactly to the one just discussed 
and leads to the same graphical output as in Figure~\,\ref{fig:fitExample}, looks
as follows: 

\begin{verbatim}
    # Python code for fitting a straight line
    from kafe2 import XYContainer, Fit, Plot
    xy_data = XYContainer(
        x_data=[1.0, 2.0, 3.0, 4.0], 
        y_data=[2.3, 4.2, 7.5, 9.4]
    )
    xy_data.add_error(
        axis='x', 
        err_val=0.05, 
        relative=True
    )
    xy_data.add_error(
        axis='y', 
        err_val=0.4
    )
    line_fit = Fit(xy_data)
    line_fit.do_fit()  
    line_fit.report()
    plot = Plot(fit_objects=line_fit) 
    plot.plot()
    plot.show()
\end{verbatim}

Since fits of models to two-dimensional data points is a very common use 
case a built-in function implementing the pipeline is also available:

\begin{verbatim}
    import kafe2
    kafe2.xy_fit(
        x_data=[1.0, 2.0, 3.0, 4.0],
        y_data=[2.3, 4.2, 7.5, 9.4],
        x_error_rel=0.05,
        y_error=0.4
    )
    kafe2.plot()
\end{verbatim}

Because relative abscissa uncertainties are specified (via \verb|x_error_rel|),
the built-in pipeline automatically switches to calculating parameter uncertainties with
the profile likelihood. A plot of the confidence intervals/regions is also created
automatically.

Users are provided with several resources to familiarize themselves with the 
use of {\it kafe2}. In addition to the traditional documentation explaining 
the use by showing examples there is also a tutorial in the form of a Jupyter 
notebook that takes a more interactive approach.
Among others, the following topics are covered:

\begin{enumerate}[label=(\alph*)]
    \item Simple linear regression with a straight line as model function;
    \item definition of arbitrary model functions;
    \item specification of uncertainties and their correlations for both 
          ordinate and abscissa values;
    \item treatment of relative uncertainties;
    \item incremental construction of covariance matrices from individual
          uncertainties;
    \item calculation of confidence intervals and contours using the profile
          likelihood method;
    \item constraining model parameters in accordance with external information;
    \item fitting distributions to binned or unbinned data using the maximum
          likelihood method;
    \item fitting models to data with non-Gaussian uncertainties, e.g.\ frequency
          distributions;
    \item simultaneous fitting of multiple models with shared parameters to
          several data sets.
\end{enumerate}

Since {\it kafe2} is open-source software, the code can be modified or extended as
needed. 

\subsection{Application example} \label{ssec_applications}
\begin{figure}[htbp] \begin{center}
\mbox{\includegraphics[width=0.9\linewidth] {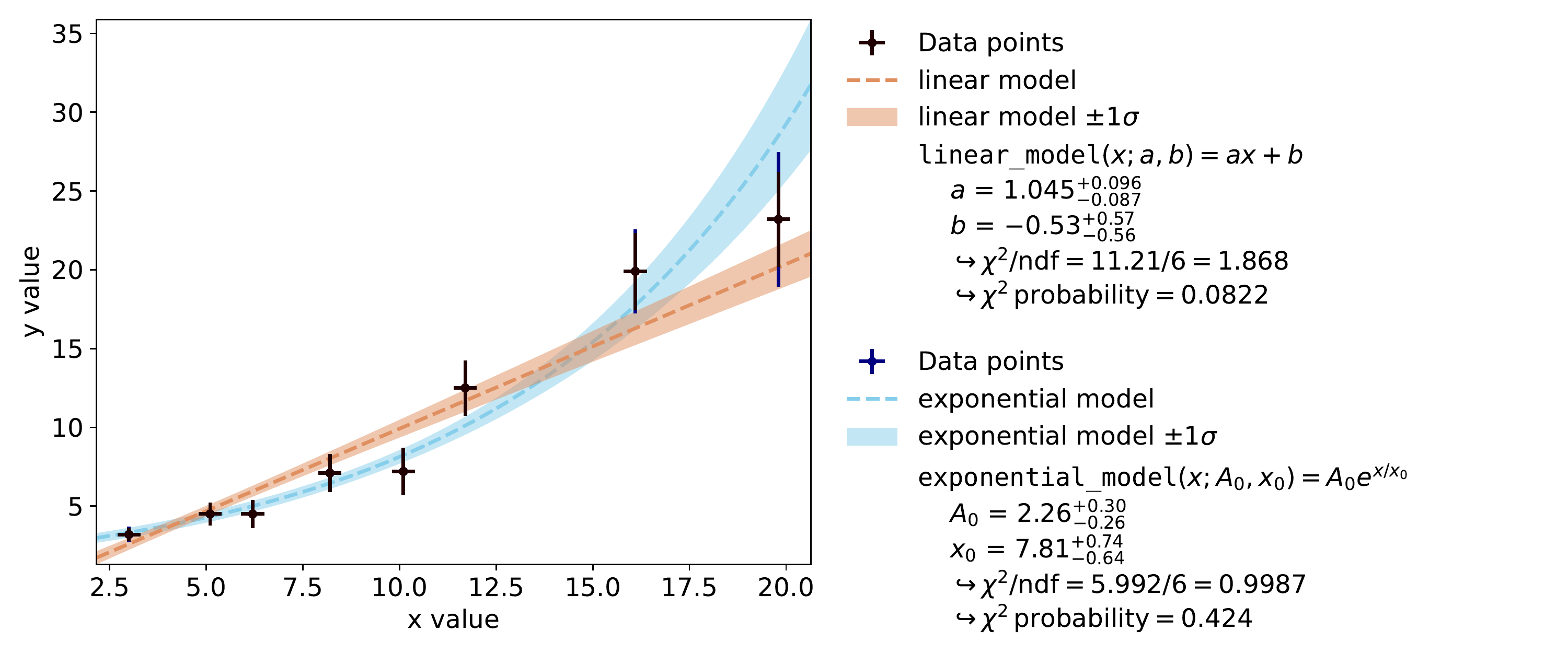}}
\vspace {-1pc} \caption[Model comparison with kafe2] 
{Fitting two models to the same data with {\it kafe2}.} 
\label{fig:twoModels} 
\end{center}\end{figure}

Figure~\,\ref{fig:twoModels} shows the results obtained when fitting both a linear and an 
exponential model to the same data.
Both the ordinate and the abscissa values are subject to measurement uncertainty. 
Additionally the uncertainties on the ordinate values are specified as relative.
Because of the reference to the model value, the uncertainties shown in the
graph differ depending on the model function.

When plotting or printing fit results, {\it kafe2} provides the user
with two metrics that can be used for hypothesis testing, specifically for Pearson's 
$\chi^2$ test. The first is the value of the cost function at the minimum divided
by the number of degrees of freedom $\mathrm{NDF}$, which has an expectation value of unity, assuming that the model adequately describes the data within its statistical fluctuations. 
The second metric quantifying the 
goodness-of-fit is the $\chi^2$ probability, i.e.\ the probability to find a 
minimum $\chi^2$ value larger than the one actually observed, with larger values indicating a better agreement between the data and the model.
Based on this value, a simple hypothesis test can be performed, rejecting the hypothesis that the data is adequately described by the model if the $\chi^2$ probability is less than a pre-defined significance threshold $\alpha$.

In the case shown here, the probability for obtaining an even higher than observed $\chi^2$ cost function value
is 8.8\% for the linear model and 44.5\% for the exponential model, indicating that both models are acceptable with a significance of $\alpha = 5\%$. As the 
visual impression already shows, the exponential model fits the data more closely.

\begin{figure}[hbtp] \begin{center}
\mbox{\includegraphics[width=0.6\linewidth] {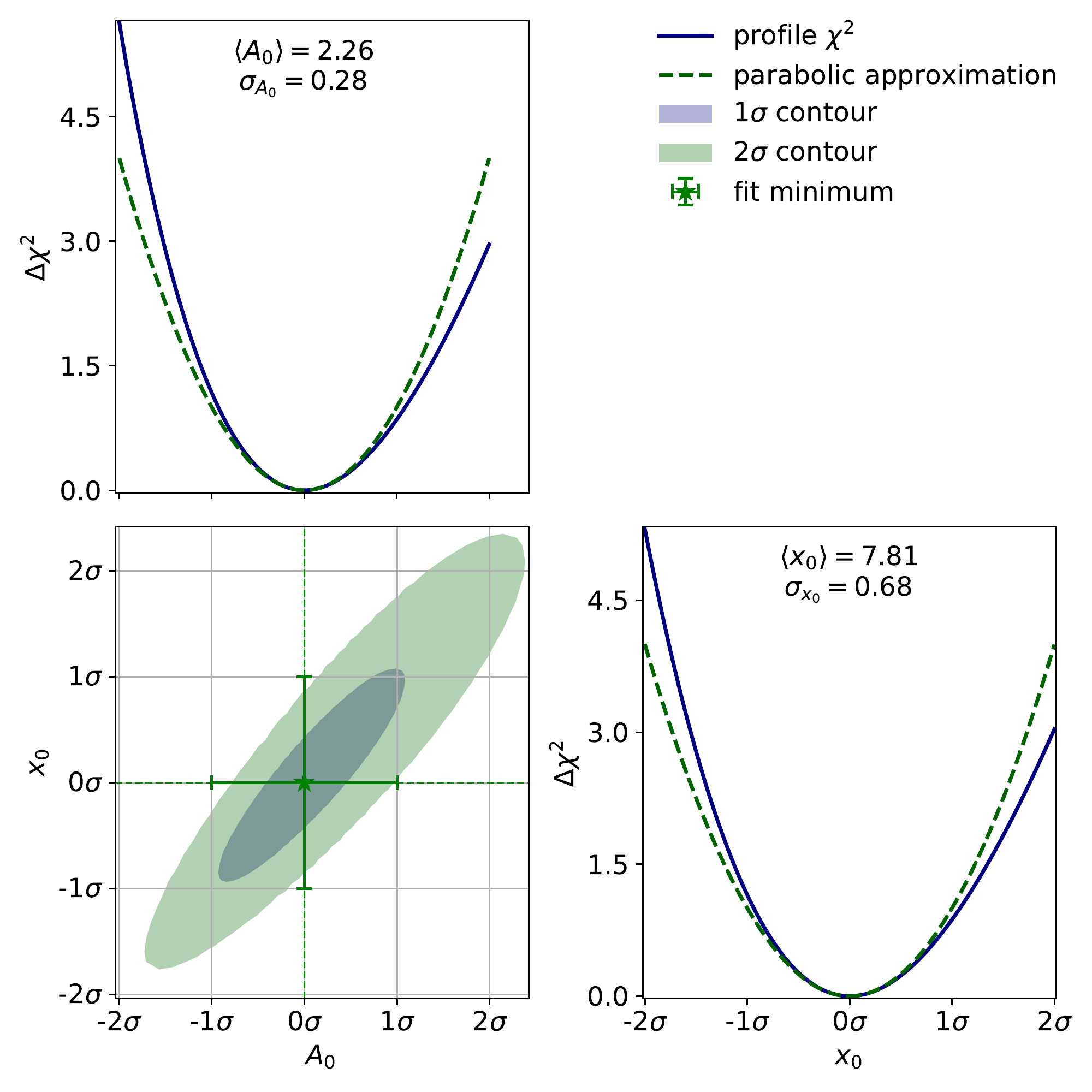}}
\vspace {-1pc} \caption[Contours and Profile Likelihood curves] 
{Two-dimensional confidence contours and profile likelihood curves for the 
exponential model.} 
\label{fig:Contours} 
\end{center}\end{figure} 

\pagebreak[3]

The parameterization of the exponential model is - intentionally - chosen 
suboptimally to amplify the nonlinearity. The parameter $x_0$ is in
the denominator of the expression in the exponent; 
if $x_0$ were replaced by a parameter $\lambda := 1/ x_0$, then the parameter
uncertainties derived from the Cramér-Rao-Fréchet bound would be more accurate.
The deviation from the parabolic shape can be inspected via the profile likelihood
curve, as shown in Figure~\ref{fig:Contours}. 
The profile likelihood for the parameter $x_0$ - drawn as a solid blue line - 
clearly deviates from the parabola that is extrapolated from the second 
derivative at the cost function minimum. The 68.3\% confidence interval is determined 
by finding the intersection of the likelihood curves with a constant function equal to 1;
the confidence interval derived from the profile likelihood method differs
from the conventional confidence interval obtained by extrapolating the second derivatives.
Instead of a symmetric uncertainty of $\pm$0.70, this results in an asymmetric confidence
interval from -0.66 to +0.76 around the central value. 

The exponential model is a linear function of the parameter $A_0$.
Despite this, the profile likelihood of $A_0$ also differs from a parabola:
the 68.3\% confidence interval ranges from -0.27 to +0.32 relative to the central value. 
This effect is mostly caused by the relative uncertainties on the ordinate values, which 
are specified as relative to the model values. As such, larger model values 
increase the uncertainties and therefore result in a lower cost function value.
Conversely, smaller model values result in a higher cost function value.
The asymmetry of the uncertainty of parameter $a$ shown as the result of the linear fit 
in Figure~\ref{fig:twoModels} is caused by the same effect.

The third graph in the bottom left corner of Figure~\ref{fig:Contours} shows the 
boundaries of the confidence regions for the shared distribution of $x_0$ and $A_0$,
as determined by the profile likelihood method.
The plot shows the confidence contours for $\Delta\chi^2 = 1$ and 
$\Delta\chi^2 = 4$, which correspond to the $1\,\sigma$ and $2\,\sigma$ uncertainty 
ranges, respectively. The slope of the contour reflects the strong correlation of 
90\% between the parameters $A_0$ and $x_0$. 
The cross indicates the uncertainties determined from the second derivatives at 
the minimum.
The deviation from the elliptical shape is visible for both contours, but it
is particularly pronounced for the $2\,\sigma$ contour. 
This is expected since the extrapolation of the second derivatives is more
accurate close to the cost function minimum.
Students should be encouraged to include the asymmetric uncertainties as well as 
the confidence contours when documenting their results.

\section{Experiences and conclusion}\label{sec_experience}

The methods upon which {\it kafe2} is built are a core part of the physics curriculum 
at the Karlsruhe Institute of Technology.
Bachelor students are required to attend a lecture on computer-based data analysis 
which includes topics such as the basics of statistics and the visualization of data.
The lecture is aimed at students in the second semester and serves as a preparation 
for the physics laboratory courses starting in the third semester.
As such, the focus of the lecture is placed on equipping students with the tools and knowledge
necessary to perform simple but modern data analyses on their own. More complex problems
such as fits with user-defined maximum-likelihood functions or fits to histogram data
arise in the context of advanced practical courses or work on projects done as part of a Bachelor's thesis.

The {\it kafe} and {\it kafe2} packages have been in continuous use at KIT in the context
of physics laboratory courses since 2015. For ease of use, sample code is provided as part
of a collection of  useful Python functions for storage, processing and visualization 
(software package \emph{PhyPraKit}\,\cite{phyprakit}), which encapsulates the complex functionality
of  {\it kafe2} and provides a simple interface sufficient for most practical needs. 
Optional tasks offer incentives to try out more complex functionality, replacing
the traditional simple error calculation with complete uncertainty models with 
covariance matrix in the model fitting. 

More in-depth lectures on data analysis that build on the foundations laid in the 
practical courses are offered to postgraduate students. Positive feedback from supervisors of 
Bachelor's theses indicates that the students experience a significant gain in competence
when it comes to the evaluation of scientific data as well as its visualization, discussion, 
and presentation in scientific contexts. The student surveys for the courses show that the 
prior knowledge in the area of digital data processing varies greatly from student to 
student. The acceptance of the tools provided, the assessment of their usefulness, and the
positive awareness of the students' own competence increase over the course of their studies.

The lecture on computer-based data analysis is deliberately placed early in the physics 
curriculum. Students can only experience a gain in knowledge from laboratory courses if 
they understand the mechanisms that the experiments are based on - this holds true both 
for the underlying physics and the statistical methods employed.
However, the early introduction of software tools also necessitates that said tools are 
(comparatively) easy to use. Most physics students were taught at most basic programming 
in school and cannot be expected to learn how to apply the complex tools used in scientific
practice. At the same time the available simple tools (some of which require no programming 
knowledge at all) do not use state-of-the-art statistical methods and thus cannot enlighten 
students when it is necessary to apply them. To brigde this gap, {\it kafe2} aims to provide
for an easy-to-use tool built on state-of-the-art statistical methods.
While the early introduction of modern analysis methods still poses some difficulty,
most students respond positively to the challenge and are ultimately able to master it well.

With the tools commonly used in basic physics education for model fitting and
visualization, an analysis as shown in the examples above is practically impossible 
to perform. With {\em kafe2}, however, students have a sufficiently simple tool at 
their disposal to carry out data analysis tasks in accordance with contemporary 
scientific standards.

\end{document}